\documentclass[a4paper,11pt]{article}
\usepackage{pos}

\newcommand{\f}[2]{\frac{#1}{#2}} 
\newcommand{\tf}[2]{{\textstyle \frac{#1}{#2}}} 
 
\newcommand{\de}{\partial} 
\newcommand{\la}{\langle} 
\newcommand{\ra}{\rangle}

\newcommand{\Oc}{{\cal O}} 
\renewcommand{\Re}{{\rm Re}}

\newcommand{\sgnepsilon}{\varepsilon} 
\newcommand{\Dmeas}{\mathcal{D}}

\title{New approach to lattice QCD at finite density: \\ 
reweighting without an overlap problem}
\ShortTitle{New approach to lattice QCD at finite density: reweighting without an overlap problem}

\author*[a]{Attila P\'asztor}
\author[b]{Szabolcs Bors\'anyi}
\author[a,b,c,d,e]{Zolt\'an Fodor} 
\author[a]{Matteo Giordano}
\author[a]{Korn\'el Kap\'as}
\author[a,f]{S\'andor D.\ Katz}
\author[a]{D\'aniel N\'ogr\'adi}
\author[b]{Chik Him Wong}

  \affiliation[a]{ELTE E\"otv\"os Lor\'and University, Institute for
  Theoretical Physics, P\'azm\'any P\'eter s\'et\'any 1/A, H-1117, Budapest,
  Hungary}
  \affiliation[b]{Department of Physics, Wuppertal University, Gaussstr.\ 20, D-42119, Wuppertal, Germany}
  \affiliation[c]{Pennsylvania State University, Department of Physics, State College, Pennsylvania 16801, USA}
  \affiliation[d]{J{\"u}lich Supercomputing Centre, Forschungszentrum J{\"u}lich, D-52425 J{\"u}lich, Germany}
  \affiliation[e]{Physics Department, UCSD, San Diego, CA 92093, USA}
  \affiliation[f]{MTA-ELTE Theoretical Physics Research Group,
  P\'azm\'any P\'eter s\'et\'any 1/A, H-1117 Budapest, Hungary.}

\emailAdd{apasztor@bodri.elte.hu}

\abstract{
Approaches to finite baryon density lattice QCD usually suffer from uncontrolled systematic uncertainties in addition to the well-known sign problem. 
We test a method - sign reweighting - that works directly at finite chemical potential and is yet free from any such uncontrolled systematics: 
with this approach the only problem is the sign problem itself. In practice the approach involves the generation of configurations with the positive fermionic 
weights given by the absolute value of the real part of the quark determinant, and a reweighting by a sign.  There are only two sectors, +1 and -1 and as long 
as the average $\left\langle \pm \right\rangle \neq 0$ (with respect to the positive weight) this discrete reweighting has no overlap problem - unlike reweighting
from $\mu=0$ - and the results are reliable. We also present results based on this algorithm on the phase diagram of lattice QCD with two different actions: as a 
first test, we apply the method to calculate the position of the critical endpoint with unimproved staggered fermions at $N_\tau=4$; as a second application, we 
study the phase diagram with 2stout improved staggered fermions at $N_\tau=6$. This second one is already a reasonably fine lattice - relevant for phenomenology.
We demonstrate that the method penetrates the region of the phase diagram where the Taylor and imaginary chemical potential methods lose predictive power.
}

\FullConference{%
 The 38th International Symposium on Lattice Field Theory, LATTICE2021
  26th-30th July, 2021
  Zoom/Gather@Massachusetts Institute of Technology
}

\begin{document}
\maketitle

\section{Introduction}
One of the most important unsolved problems in QCD is the 
determination of the phase diagram of strongly interacting matter in the 
temperature-baryochemical potential plane. The most well-established method 
for first-principles studies of QCD is the lattice. Finite chemical potential 
lattice calculations are, however, hampered by the notorious sign- or 
complex-action problem. A number of approaches have been proposed to side-step this 
problem, such as reweighting~\cite{Hasenfratz:1991ax,Fodor:2001au,Fodor:2004nz,Fodor:2007vv,Endrodi:2018zda,Giordano:2019gev,Giordano:2020uvk,Giordano:2020roi,Borsanyi:2021hbk}, Taylor expansion around zero 
chemical potential~\cite{Allton:2002zi,Gavai:2008zr,Bellwied:2015lba,Bazavov:2017dus,HotQCD:2018pds,Giordano:2019slo}, 
analytic continuation from purely imaginary chemical potentials~\cite{deForcrand:2002hgr,DElia:2002tig,Cea:2015cya,DElia:2016jqh,Bonati:2018nut,Bellwied:2019pxh,Borsanyi:2020fev,Pasztor:2020dur,Bellwied:2021nrt,Borsanyi:2021sxv}, or the 
complex Langevin approach~\cite{Parisi:1983mgm, Aarts:2009uq, Sexty:2013ica}. Unfortunately, all of these methods
introduce extra problems, which are different from the sign problem itself,
such as the analytic continuation problem of the Taylor and imaginary chemical potential
methods, the overlap problem of reweighting and the Taylor method, or the convergence issues of 
complex Langevin. Often these problems are just as prohibiting as the original sign problem. 
It is therefore of value to have an alternative method, which does not have any of the 
extra problems of the existing methods: a method where the only problem is the 
sign problem itself.

Although manifesting as different, the analytic 
continuation problem of the Taylor and imaginary chemical potential
methods and the overlap problem of the reweighting and Taylor methods 
have the same physical origin: an inability to directly sample the gauge 
configurations most relevant to finite chemical potential, thus requiring 
some kind of extrapolation in both cases. One
would then  like to perform simulations in a theory from which
reconstruction of the desired theory is the least affected by such
systematic effects, by (1) keeping as close as possible to the most relevant configurations, 
thus minimizing the overlap problem, and by (2) making the complex-action problem, or 
sign problem, due to cancellations among contributions, as mild as
possible. We show here that such an approach - sign reweighting - has already 
become feasible on phenomenologically relevant lattices. This conference contribution
is mainly based on Refs.~\cite{Giordano:2020roi,Borsanyi:2021hbk}

\section{Reweighting and the overlap problem}

A generic reweighting method reconstructs expectation values in a
desired target theory, with microscopic variables $U$, path-integral
weights $w_t(U)$, and partition function
$Z_t = \int\Dmeas U \ w_t(U)$, using simulations in a theory with real
and positive path-integral weights $w_s(U)$ and partition function
$Z_s = \int\Dmeas U \ w_s(U)$, via the formula:
\begin{equation}
  \label{eq:reweight}
  \left\langle \Oc \right\rangle_t = \frac{\left\langle \frac{w_t}{w_s}
      \Oc \right\rangle_s }{\left\langle
      \frac{w_t}{w_s}\right\rangle_s}\,
\end{equation}
where $\langle \dots \rangle_t$ and $\langle \dots \rangle_s$ means expectation
value in the target and the simulated theories respectively.
When the target theory is lattice QCD at finite chemical potential,
the target weights $w_t(U)$ have wildly fluctuating phases: this is
the infamous sign problem. In addition 
to this problem, generic reweighting methods also suffer from an overlap 
problem: the probability distribution of the reweighting factor $w_t/w_s$ has
generally a long tail, which cannot be sampled efficiently
in standard Monte Carlo simulations. The overlap problem is present even in 
cases when one tries to reweight to a theory without a sign problem, such 
as reweighting to a different bare gauge coupling. However, even in the case of 
reweighting from zero to finite baryochemical potential, 
it is actually the overlap problem, rather than the sign problem, that
constitutes the immediate bottleneck in QCD when one tries to extend
reweighting results to finer lattices, even in the case of the multiparameter
reweighting method~\cite{Fodor:2004nz}, as was recently demonstrated 
by the study of the histogram of the reweighting factors $w_t/w_s$ 
in Ref.~\cite{Giordano:2020uvk}.

A way to address the overlap problem is to reweight from a theory where the
reweighting factors $w_t/w_s$ takes values in a compact space. In such a 
case, their distribution does not have tails by construction, and so that the
ratio $\f{Z_t}{Z_s} = \left\langle \f{w_t}{w_s} \right\rangle_s$
can be calculated without encountering any heavy-tailed distributions. 
The most obvious choice for the theory to reweight from is the phase quenched (PQ)
theory, defined by
\begin{equation}
    w_{PQ} = |\operatorname{det} M_{ud}(\mu)^{\frac{1}{2}}| \operatorname{det} M_s(0)^{\frac{1}{4}} e^{-S_g}\rm{,} 
\end{equation}
where $\mu$ is the chemical potential of the light quarks, and for simplicity we take the strange quark 
chemical potential to be zero. 
In this case the reweighting factors are pure phases $e^{i \theta}$, 
where $\theta = \operatorname{Arg} \left( \det M_{ud}(\mu)^{1/2} \right)$. 
Another choice, which - as we will show - has a weaker sign problem
is to reweight from the sign quenched (SQ) ensemble~\cite{deForcrand:2002pa,Alexandru:2005ix,Giordano:2020roi}, defined by
\begin{equation}
    w_{SQ} = |\operatorname{Re} \operatorname{det} M_{ud}(\mu)^{\frac{1}{2}}| \operatorname{det} M_s(0)^{\frac{1}{4}} e^{-S_g}\rm{,} 
\end{equation}
where reweighting involves only a sign factor:
\begin{equation}
    \sgnepsilon = \operatorname{sign} \cos \operatorname{Arg}  \operatorname{det} M_{ud}(\mu)^{\frac{1}{2}}\rm{.}
\end{equation}
We note that this amounts to the substitution of the determinant with its 
real part in the path integral, which is 
not permitted in arbitrary expectation values, but is 
completely valid for (1) observables that satisfy $O(U)=O(U^*)$ 
or (2) observables which can be defined as real derivatives of the partition function with respect to a real parameter, such as 
the gauge coupling, the quark mass or the chemical potential. While this set of observables is not exhaustive, it 
is enough to study bulk thermodynamics, which is the main target of our work.

\section{Lattice simulations}
As a first test of the method, in Ref.~\cite{Giordano:2020roi} we used unimproved 
staggered fermions on lattices of size 
$6^3 \times 4, 8^3 \times 4, 10^3 \times 4$ and $12^3 \times 4$ to estimate 
the position of the critical endpoint on these coarse lattices. For each lattice size and
each chemical potential in the baryochemical potential range $0 \leq \hat{\mu}_B = \mu_B/T \leq 2.4$, we simulated one 
single value of the bare gauge coupling $\beta$, chosen to be close to the transition temperature
at the given $\mu_B$.

To move in the direction of the continuum limit, in Ref.~\cite{Borsanyi:2021hbk} we used a tree level Symanzik improved gauge action with
two steps of stout smearing with parameter $\rho=0.15$ on $16^3\times 6$ lattices~\cite{Borsanyi:2010bp}.
We performed a scan in chemical potential
at fixed $T=140\,{\rm MeV}$, and a scan in temperature at fixed
$\hat{\mu}_B= 1.5$.  

In both cases, simulations were performed by modifying the RHMC
algorithm at $\mu_B=0$ by including an extra accept/reject step that
takes into account the factor $\f{|\Re \det M_{ud}(\mu)^{\frac{1}{2}}|}{\det M_{ud}(0)}$.  The
determinant was calculated with the reduced matrix
formalism~\cite{Hasenfratz:1991ax} and dense linear algebra, with no
stochastic estimators involved. 

\section{Comparison with phase reweighting and the strength of the sign problem}

In the PQ ensemble the severity of the sign
problem is measured by the average phase factor
$ \la e^{i\theta} \ra^{\rm PQ}_{T,\mu}= \la \cos \theta \ra^{\rm
  PQ}_{T,\mu}$, while in the SQ ensemble it is measured by the average sign
$\la \sgnepsilon\ra^{\rm SQ}_{T,\mu}= {\la\cos \theta\ra^{\rm
    PQ}}/{\la |\!\cos \theta| \ra^{\rm PQ}}$.  Clearly,
$\la \cos \theta \ra^{\rm PQ}_{T,\mu} \le \la \sgnepsilon \ra^{\rm
  SQ}_{T,\mu}$, 
so the sign problem is generally weaker in the SQ case. To understand how much 
weaker it is, it is useful to introduce some small chemical potential approximations
for the strength of the sign problem in both cases.
The probability distribution of the phases $\theta= \arg \det M$ in the
phase quenched theory, $P_{\rm PQ}(\theta)$, controls the strength of
the sign problem in both ensembles.  A simple estimate
can then be obtained with the following two steps:
(\textit{i}) in a leading order cumulant expansion,
$P_{\rm PQ}(\theta)$ is assumed to be a wrapped Gaussian distribution;
(\textit{ii}) the chemical potential dependence of its width is
approximated by 
the leading order term in its Taylor expansion~\cite{Allton:2002zi},
\begin{equation}
\sigma(\mu)^2 \approx \left\langle \theta^2 \right\rangle_{\rm LO}
    =-\frac{4}{9} \chi^{ud}_{11} \left( LT \right)^3\hat{\mu}_B^2 \rm{,}
\end{equation}
where
$\chi^{ud}_{11} = \f{1}{T^2}\frac{\partial^2 p }{\partial
  \mu_u \partial \mu_d
  }|_{\mu_u=\mu_d=0}$
is the disconnected part of the light
quark susceptibility, obtained in $\mu=0$ simulations.  In this
approximation 
the strength of the sign problem can be calculated 
analytically in both cases, with
$\la \cos \theta \ra^{\rm PQ}_{T,\mu} \approx
e^{-\frac{\sigma^2(\mu)}{2}}$ in the phase quenched case, while in the
sign quenched case the expression for
$\la \sgnepsilon\ra^{\rm SQ}_{T,\mu}$ is more involved. (See the Appendix of Ref.~\cite{Borsanyi:2021hbk}.) 
\begin{figure*}[t]
  \centering
  \includegraphics[width=0.45\textwidth]{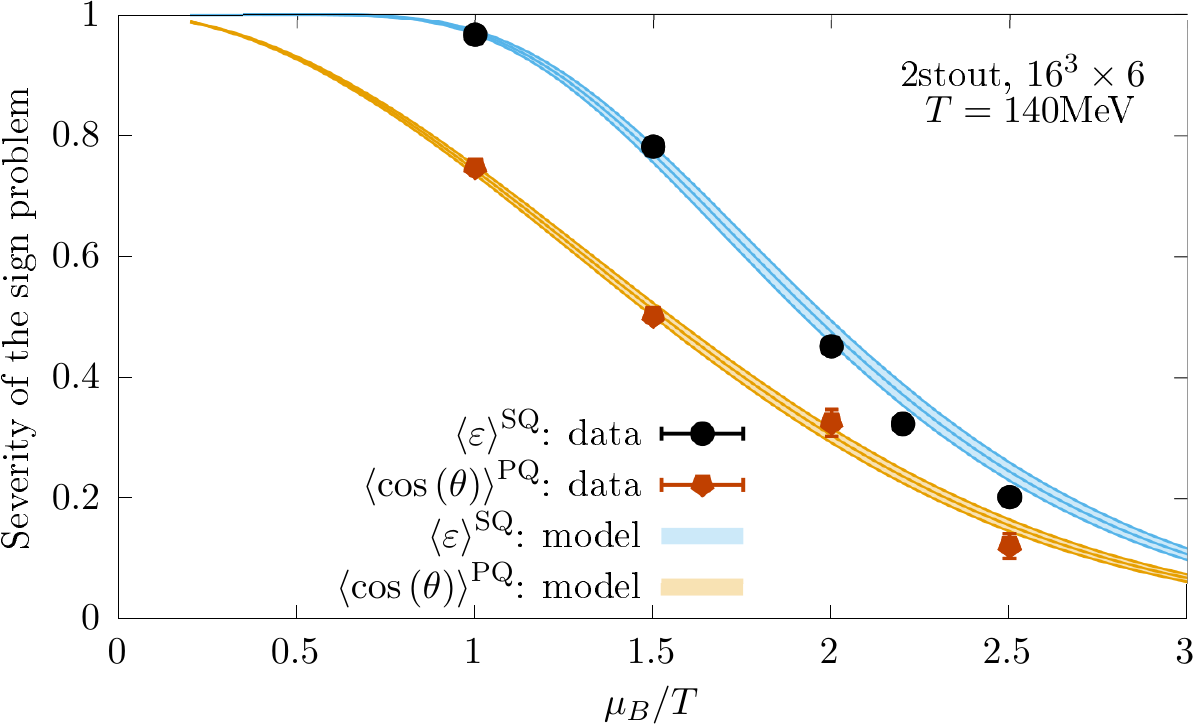}
  \includegraphics[width=0.45\textwidth]{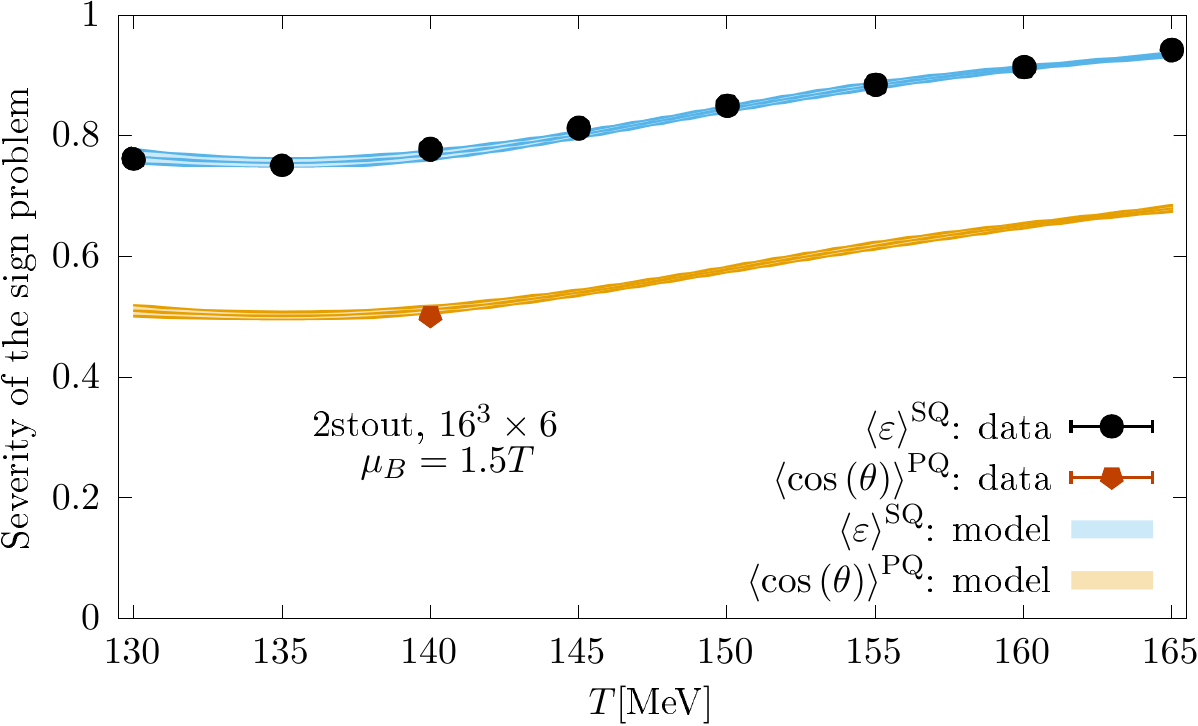}
  \caption{The strength of the sign problem on 2stout improved $16^3 \times 6$ staggered lattices 
    as a function of $\mu_B/T$
    at $T=140$ MeV (left) and as a function of $T$ at $\mu_B/T=1.5$. A
    value close to $1$ shows a mild %weak
    sign problem, while a small value
    indicates a severe sign problem.  Data for sign reweighting
    (black) and phase reweighting (orange) are from
    simulations.  Predictions of the Gaussian model (see text) are also shown.}
  \label{fig:sign}
\end{figure*}
It is however worth noting the different asymptotics of the two cases.  The small-$\mu$
(i.e., small-$\sigma$) asymptotics are notably very different, with
$ \la \cos\theta\ra^{\rm PQ}_{T,\mu} \sim 1 - \f{\sigma^2(\mu)}{2}$ analytic
in $\hat{\mu}_B$,
while in the sign quenched case
$\la \sgnepsilon\ra^{\rm SQ}_{T,\mu}$ is not analytic,
\begin{equation}
  \la \sgnepsilon\ra^{\rm SQ}_{T,\mu}
    \underset{\hat{\mu}_B\to  0}{\sim}
  1 -\left(\tf{4}{\pi} \right)^{\f{5}{2}}
  \left(\tf{\sigma^2(\mu)}{2}\right)^{\f{3}{2}} e^{-\f{\pi^2}{8
      \sigma^2(\mu)}}\,,
\end{equation}
approaching 1 faster than any polynomial: therefore we expect the sign problem of the 
phase quenched ensemble to get worse faster at small $\mu_B$ as compared to the sign quenched
case. The large-$\mu$ or large volume
asymptotics are on the other hand quite similar: in the large-$\sigma$ limit
$P_{PQ}(\theta)$ tends to the uniform distribution, and so one arrives at
$  \f{\la \sgnepsilon\ra^{\rm SQ}_{T,\mu}}{\la \cos\theta\ra^{\rm PQ}_{T,\mu}}
  \underset{\hat{\mu}_B\,\text{or}\, V\to \infty}{\sim}
  \left(
    \int_{-\pi}^{\pi}d\theta\, |\!\cos \theta| \right)^{-1}=\frac{\pi}{2} \,, $
which asymptotically translates to a factor of
$(\f{\pi}{2})^2 \approx 2.5$ less statistics needed for a sign
quenched simulation as compared to a phase quenched simulation.

The simple considerations made above
are confirmed by actual simulation data to a decent degree,
as can be seen in Fig.~\ref{fig:sign}: our simple model predicts the strength of the sign problem
both as a function of $\mu_B$ at a fixed temperature (left) and as a function of temperature at a 
fixed $\mu_B/T$ (right). This is of great practical 
importance, as it makes the planning of future simulation projects
with reweighting relatively straightforward.

\section{Physics observables}
\subsection{Unimproved action at $N_\tau=4$}

\begin{figure*}[t]
  \centering
  \includegraphics[width=0.45\textwidth]{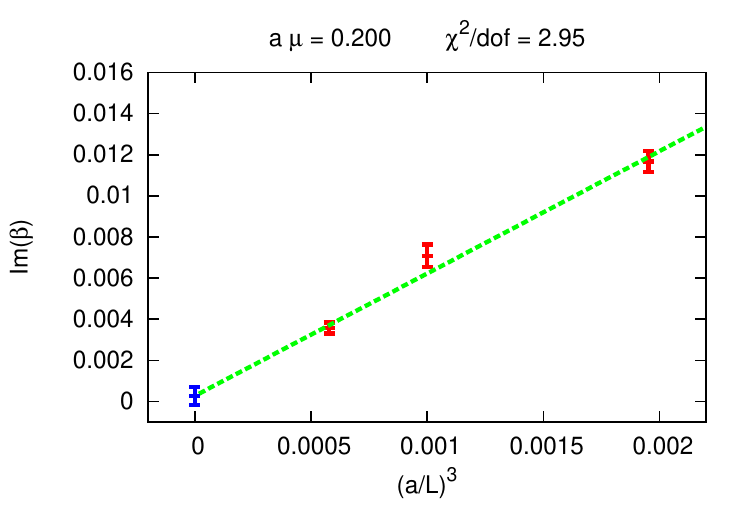}
  \includegraphics[width=0.45\textwidth]{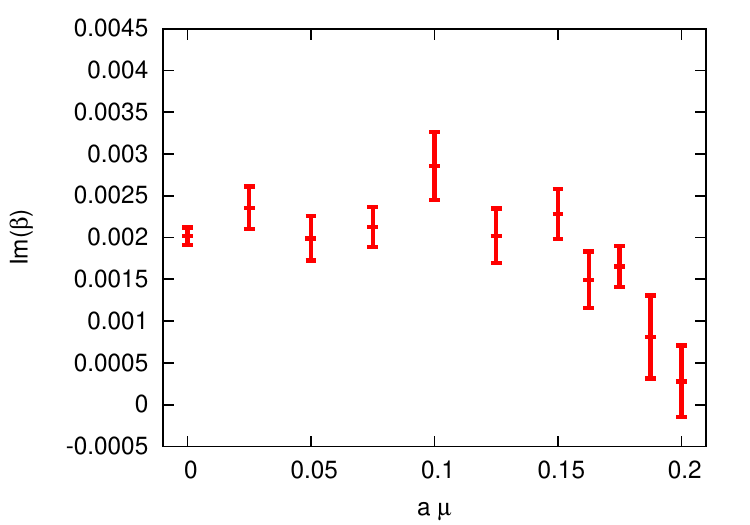}
    \caption{Finite volume scaling of the imaginary part of the leading Fisher zero at $\mu_B/T = 2.4$ (left) and the infinite volume extrapolated value of the 
    imaginary part of the leading Fisher zero as a function of the chemical potential (right) for the unimproved staggered action at $N_\tau=4$.
  \label{fig:fisher_unimp}
    }
\end{figure*}

The main goal of our first numerical study in Ref.~\cite{Giordano:2020roi} was to confirm or falsify the critical endpoint
prediction of Ref.~\cite{Fodor:2004nz} for the unimproved staggered discretization at temporal extent $N_\tau=4$, with a method that does 
not suffer from an overlap problem in the reweighting factors. To be as close
to Ref.~\cite{Fodor:2004nz} as possible, we 
therefore uses the same physical observable and computed the zeros of the partition function in the bare gauge coupling $\beta$, the so-called Fisher zeros.
This amounts to measuring the observables $O_\beta(U) = e^{-\frac{\beta-\beta_s}{\beta_s}S_g(U)}$, 
where $S_g(U)$ is the gauge 
action and $\beta_s$ is the simulated bare coupling. Since $O_\beta(U^{*}) = O_\beta(U)$, sign reweighting can be applied to this observable. 
The partition function has several zeros as a function of complex $\beta$. We computed 
the one closest to the real axis, which in every run 
happens to coincide with the one closest to the simulation point in the $\beta$ as well. This zero will be called the leading
Fisher zero. The finite volume scaling and the infinite volume extrapolated leading zero position can be seen in Fig.~\ref{fig:fisher_unimp}. 
As can be seen in that Figure, the leading Fisher zero extrapolates to a point on 
the real line at around $\mu_B/T=2.4$, in agreement
with the result of Ref.~\cite{Fodor:2004nz}.

\subsection{Two-stout improved action at $N_\tau=6$}

We now proceed to display physics results for the light quark condensate and density
from simulations with the 2stout improved staggered action at $N_\tau=6$.  
The light-quark chiral condensate was obtained via the formula
\begin{equation}
  \label{eq:chiralcondensate}
  \begin{aligned}
    \la\bar{\psi}\psi\ra_{T,\mu} &= \f{1}{Z(T,\mu)}\f{\de
      Z(T,\mu)}{\de
      m_{ud}} 
    =\f{T}{V}\f{1}{\la\sgnepsilon\ra^{\rm SQ}_{T,\mu}}\left\la
      \sgnepsilon\f{\de}{\de m_{\rm ud}} \ln\left |\Re \det M_{ud}^{\f{1}{2}}\right|
    \right\ra^{\rm SQ}_{T,\mu} \,,
  \end{aligned}
\end{equation}
with the determinant $\det M=\det M(U,m_{ud},m_s,\mu)$ calculated in
the reduced matrix formalism at different light-quark masses and fed
into a symmetric difference,
$\f{df(m)}{dm} \approx \f{f(m+\Delta m)-f(m-\Delta m)}{2\Delta m}$.
The step $\Delta m$ small enough to make the systematic error from the
finite difference negligible compared to the statistical error.  The
renormalized condensate was obtained with the prescription
\begin{equation}
  \label{eq:chiralcondensate3}
  \la \bar{\psi}\psi\ra_R(T,\mu) =   -\f{m_{ud}}{f_\pi^4}\left[
    \la \bar{\psi}\psi\ra_{T,\mu} -\la \bar{\psi}\psi\ra_{0,0}
  \right]\,.
\end{equation}
We also calculated the light quark density 
\begin{equation}
\begin{aligned}
  \chi^l_1 &\equiv \f{\partial \left(p/T^4\right)}{\partial \left(
      \mu/T \right) } = \frac{1}{VT^3} \f{1}{Z(T,\mu)} \f{\partial
    Z(T,\mu)}{\partial \hat{\mu}}  %\,.
  = \f{1}{VT^3 \la\sgnepsilon\ra^{\rm SQ}_{T,\mu}} \left\la
    \sgnepsilon\f{\de}{\de \hat{\mu}} \ln\left |\Re \det M_{ud}^{\f{1}{2}}\right|
  \right\ra^{\rm SQ}_{T,\mu} \,,
\end{aligned}
\end{equation}
evaluating the derivative analytically using the reduced matrix
formalism. This quantity does not have to be renormalized.

\begin{figure*}[t]
  \centering
  \includegraphics[width=0.45\textwidth]{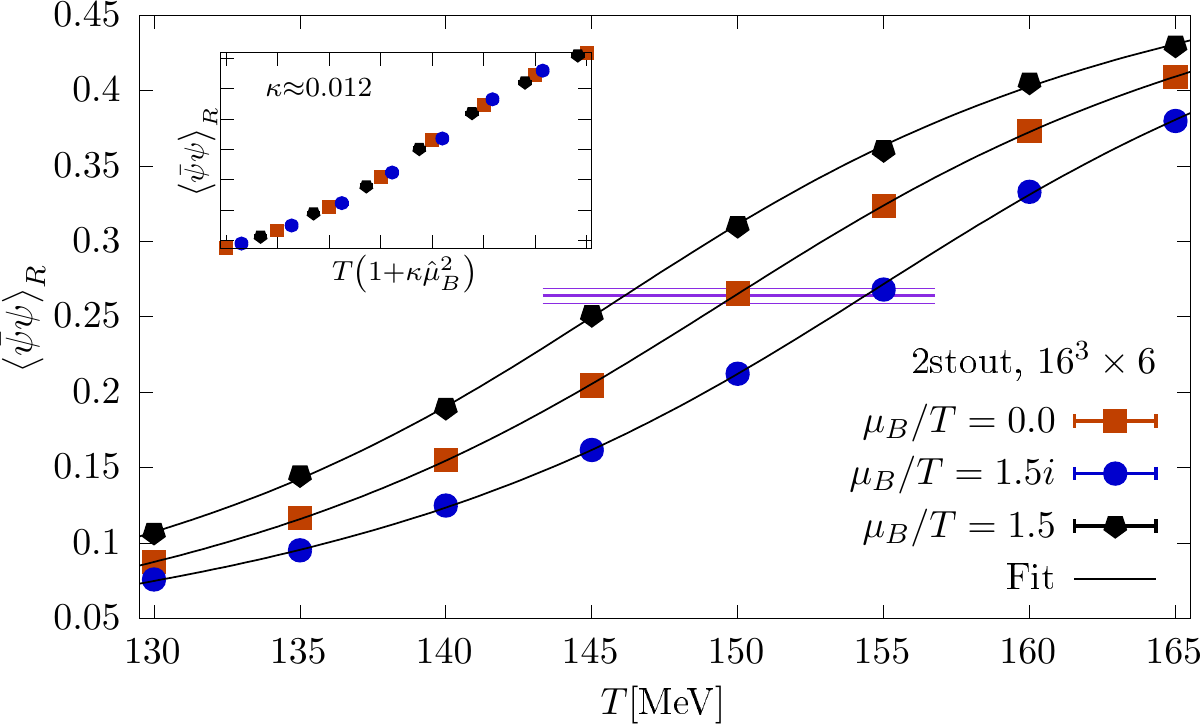}
  \includegraphics[width=0.45\textwidth]{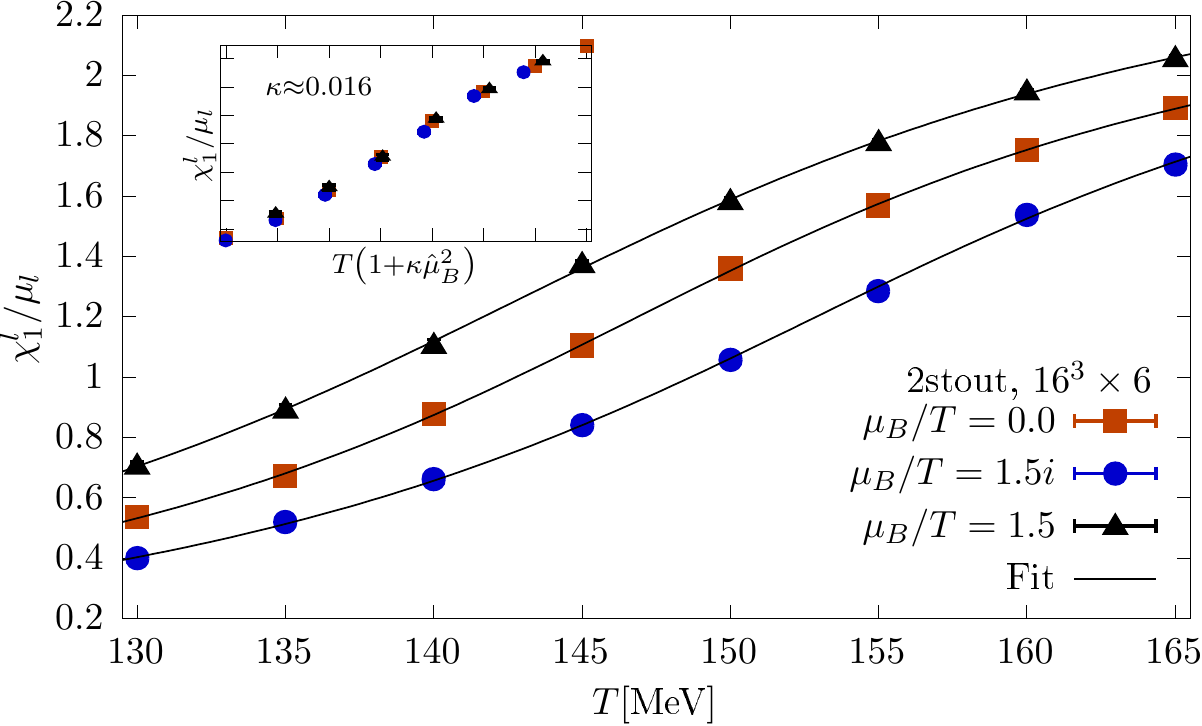}
  \caption{The renormalized chiral condensate (left) and the light
    quark number-to-light quark chemical potential ratio (right) as a
    function of $T$ at fixed $\mu_B/T=1.5,0$ and $1.5 i$ on 2stout improved 
    lattices at $N_\tau=6$. The insets show a rescaling of the temperature axis,
    which approximately collapses the curves onto each other. 
    }
  \label{fig:Tscan}
\end{figure*}

\begin{figure*}[t]
  \centering
  \includegraphics[width=0.45\textwidth]{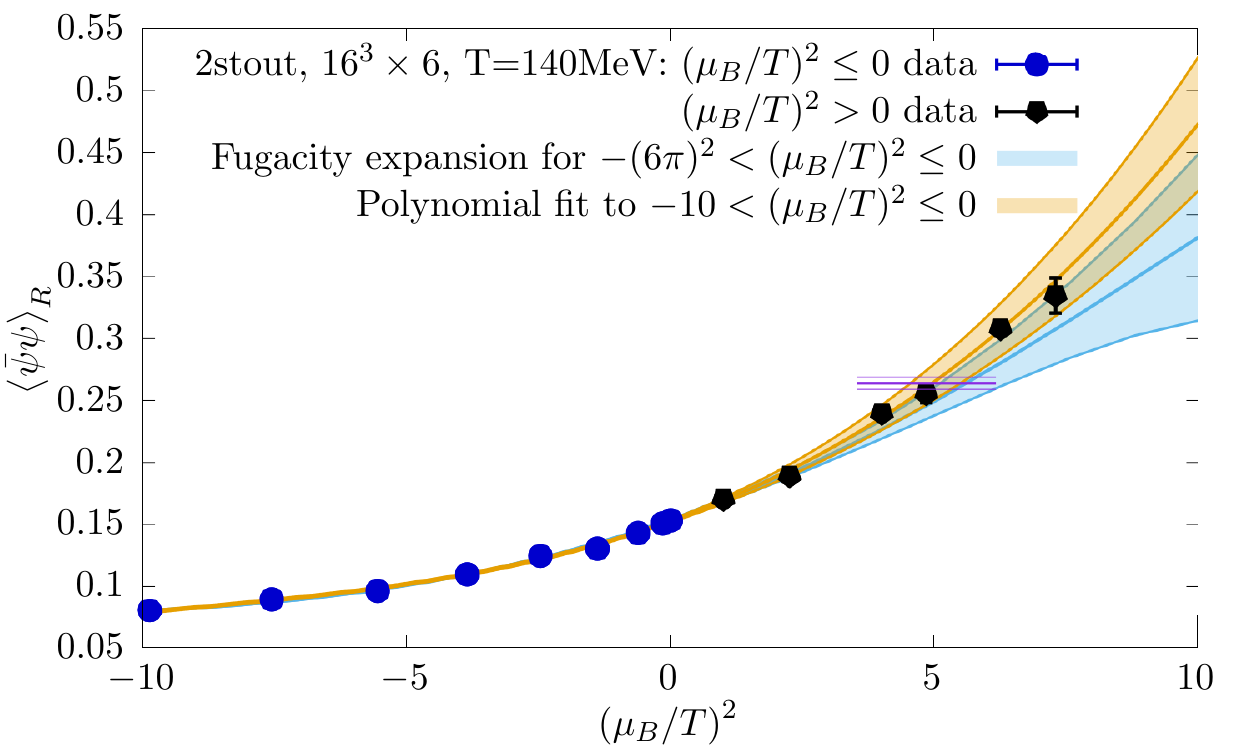}
  \includegraphics[width=0.45\textwidth]{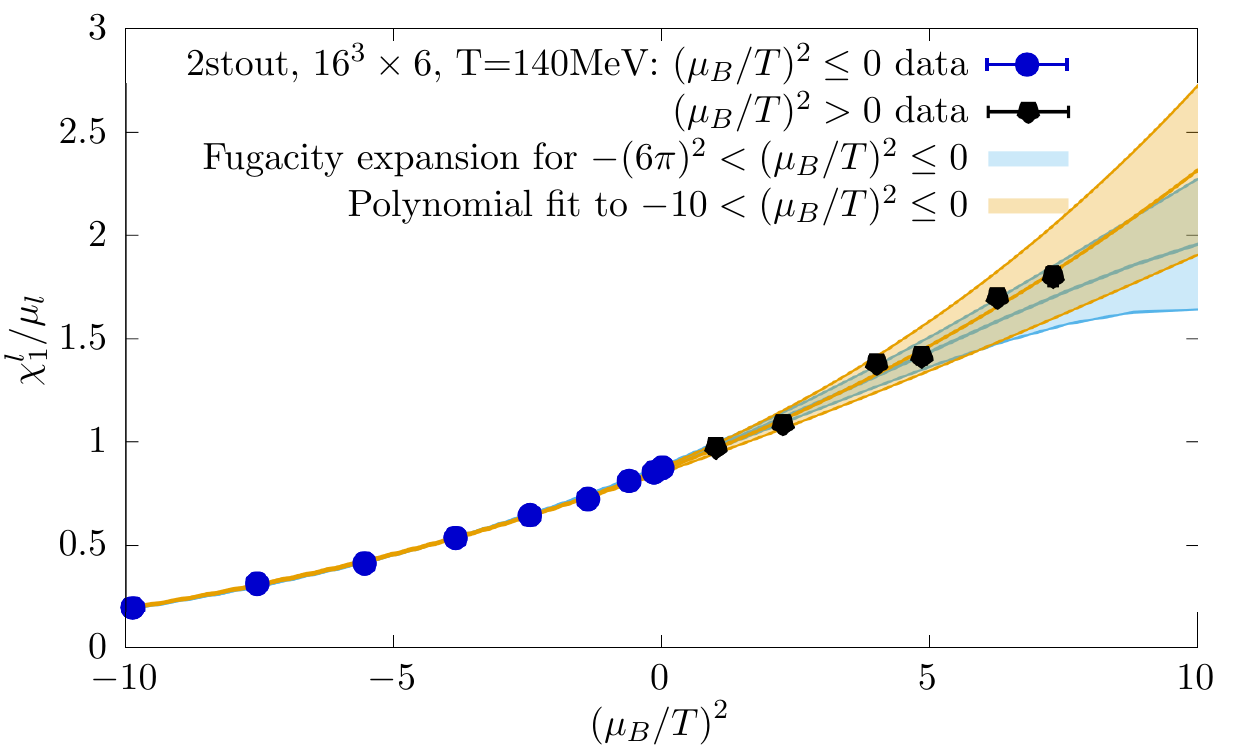}
  \caption{The renormalized chiral condensate (left) and the light
    quark number-to-light quark chemical potential ratio (right) as a
    function of $\left( \mu_B/T \right)^2$ at temperature $T=140$
    MeV with the 2stout improved staggered action at $N_\tau=6$. Data from simulations at real $\mu_B$ (black) are compared
with analytic continuation from simulations at imaginary $\mu_B$ (blue).
    In the
    left panel the value of the condensate at the crossover
    temperature at $\mu_B=0$ is also shown by the horizontal line.  
    The simulation data cross
    this line at $\mu_B/T \approx 2.2$.
    }
  \label{fig:muscan}
\end{figure*}

Our results for the temperature scan at $\mu_B/T=1.5$ are shown in 
Fig.~\ref{fig:Tscan}. We also show the corresponding curves at zero and 
the imaginary value of $\mu_B/T =1.5i$ for comparison. We also show a 
rescaling of the temperature axis, collapsing the curves into each other,
and demonstrating that at least up to $\mu_B/T=1.5$ the chiral crossover
does not get narrower. Our results for the chemical potential scan at a fixed temperature of
$T=140$ MeV are shown in Fig.~\ref{fig:muscan}.  We have performed
simulations at 
$\hat{\mu}_B=1, 1.5, 2, 2.2, 2.5, 2.7$.
The point at $\hat{\mu}_B=2.2$
corresponds roughly to the chiral transition, as at this
point the chiral condensate is close to its value at the $\mu_B=0$
crossover. The sign-quenched results are compared with the results of 
analytic continuation from imaginary chemical potentials.
To demonstrate the magnitude of the systematic errors of such an 
extrapolation we considered two fits.  (\textit{i}) As the simplest
ansatz, we fitted the data with a cubic polynomial in
$\hat{\mu}_B^2$
in the range
$\hat{\mu}_B^2 \in [-10,0]$.
(\textit{ii}) As an alternative, we
also used suitable ans\"atze for
$\left\langle \bar{\psi}\psi\right\rangle_R$ and
$\chi^l_1/\hat{\mu}_l$ based on the fugacity expansion
$p/T^4 = \sum_n A_n \cosh(n \hat{\mu})$, fitting the data in the
entire imaginary-potential range
$\hat{\mu}_B^2 \in \left[ -(6 \pi)^2,0\right]$
using respectively 7 and
6 fitting parameters.  Fit results are also shown in
Fig.~\ref{fig:muscan}; only statistical errors are displayed.  While
sign reweighting and analytic continuation 
give compatible results, in the upper half of the
$\mu_B$ range the errors from sign reweighting are an order of
magnitude smaller.  In fact, sign reweighting can penetrate the region
$\hat{\mu}_B>2$ where the extrapolation of many quantities is not yet
possible with standard methods~\cite{Bazavov:2017dus,Borsanyi:2020fev}.

\section{Summary}
Sign reweighting has opened up a new window to study the bulk thermodynamics of strongly interacting matter from first principles.
While the method is ultimately bottlenecked by the sign problem, in the region of its applicability it offers excellent reliability
compared to the dominant methods of Taylor expansion and imaginary chemical potentials - which always provide results having a shadow of a doubt
hanging over them due to the analytic continuation problem. We have demonstrated that the strength of the sign problem can be easily estimated
with $\mu=0$ simulations, making the method practical and the planning of simulation projects straightforward. We have also demonstrated that
the method extends well into the regime where the established methods start to lose predictive power and covers the range of the RHIC BES. 

\section*{Acknowledgements}
The project was supported by the BMBF Grant No. 05P18PXFCA.
This work was also supported by the Hungarian National Research,
Development and Innovation Office, NKFIH grant KKP126769.
A.P. is supported by the J. Bolyai Research
Scholarship of the Hungarian Academy of Sciences and by the \'UNKP-21-5 New
National Excellence Program of the Ministry for Innovation and Technology
from the source of the National Research, Development and Innovation Fund.
The authors gratefully acknowledge the Gauss Centre for Supercomputing e.V.
(www.gauss-centre.eu) for funding this project by providing computing time on the
GCS Supercomputers JUWELS/Booster and JURECA/Booster at FZ-Juelich.


\begin{thebibliography}{99}

%\cite{Hasenfratz:1991ax}
\bibitem{Hasenfratz:1991ax}
A.~Hasenfratz and D.~Toussaint,
%``Canonical ensembles and nonzero density quantum chromodynamics,''
Nucl. Phys. B \textbf{371} (1992), 539-549
doi:10.1016/0550-3213(92)90247-9
%155 citations counted in INSPIRE as of 28 Nov 2021

%\cite{Fodor:2001au}
\bibitem{Fodor:2001au}
Z.~Fodor and S.~D.~Katz,
%``A New method to study lattice QCD at finite temperature and chemical potential,''
Phys. Lett. B \textbf{534} (2002), 87-92
doi:10.1016/S0370-2693(02)01583-6
[arXiv:hep-lat/0104001 [hep-lat]].
%567 citations counted in INSPIRE as of 28 Nov 2021

%\cite{Fodor:2004nz}
\bibitem{Fodor:2004nz}
Z.~Fodor and S.~D.~Katz,
%``Critical point of QCD at finite T and mu, lattice results for physical quark masses,''
JHEP \textbf{04} (2004), 050
doi:10.1088/1126-6708/2004/04/050
[arXiv:hep-lat/0402006 [hep-lat]].
%951 citations counted in INSPIRE as of 27 Oct 2021

%\cite{Fodor:2007vv}
\bibitem{Fodor:2007vv}
Z.~Fodor, S.~D.~Katz and C.~Schmidt,
%``The Density of states method at non-zero chemical potential,''
JHEP \textbf{03} (2007), 121
doi:10.1088/1126-6708/2007/03/121
[arXiv:hep-lat/0701022 [hep-lat]].
%146 citations counted in INSPIRE as of 28 Nov 2021

%\cite{Endrodi:2018zda}
\bibitem{Endrodi:2018zda}
G.~Endrodi, Z.~Fodor, S.~D.~Katz, D.~Sexty, K.~K.~Szabo and C.~Torok,
%``Applying constrained simulations for low temperature lattice QCD at finite baryon chemical potential,''
Phys. Rev. D \textbf{98} (2018) no.7, 074508
doi:10.1103/PhysRevD.98.074508
[arXiv:1807.08326 [hep-lat]].
%7 citations counted in INSPIRE as of 28 Nov 2021

%\cite{Giordano:2019gev}
\bibitem{Giordano:2019gev}
M.~Giordano, K.~Kapas, S.~D.~Katz, D.~Nogradi and A.~Pasztor,
%``Radius of convergence in lattice QCD at finite $\mu_B$ with rooted staggered fermions,''
Phys. Rev. D \textbf{101} (2020) no.7, 074511
doi:10.1103/PhysRevD.101.074511
[arXiv:1911.00043 [hep-lat]].
%17 citations counted in INSPIRE as of 27 Oct 2021

%\cite{Giordano:2020uvk}
\bibitem{Giordano:2020uvk}
M.~Giordano, K.~Kapas, S.~D.~Katz, D.~Nogradi and A.~Pasztor,
%``Effect of stout smearing on the phase diagram from multiparameter reweighting in lattice QCD,''
Phys. Rev. D \textbf{102} (2020) no.3, 034503
doi:10.1103/PhysRevD.102.034503
[arXiv:2003.04355 [hep-lat]].
%5 citations counted in INSPIRE as of 28 Nov 2021

%\cite{Giordano:2020roi}
\bibitem{Giordano:2020roi}
M.~Giordano, K.~Kapas, S.~D.~Katz, D.~Nogradi and A.~Pasztor,
%``New approach to lattice QCD at finite density; results for the critical end point on coarse lattices,''
JHEP \textbf{05} (2020), 088
doi:10.1007/JHEP05(2020)088
[arXiv:2004.10800 [hep-lat]].
%10 citations counted in INSPIRE as of 27 Oct 2021

%\cite{Borsanyi:2021hbk}
\bibitem{Borsanyi:2021hbk}
S.~Borsanyi, Z.~Fodor, M.~Giordano, S.~D.~Katz, D.~Nogradi, A.~Pasztor and C.~H.~Wong,
%``Lattice simulations of the QCD chiral transition at real baryon density,''
[arXiv:2108.09213 [hep-lat]].
%3 citations counted in INSPIRE as of 27 Oct 2021

%\cite{Allton:2002zi}
\bibitem{Allton:2002zi}
C.~R.~Allton, S.~Ejiri, S.~J.~Hands, O.~Kaczmarek, F.~Karsch, E.~Laermann, C.~Schmidt and L.~Scorzato,
%``The QCD thermal phase transition in the presence of a small chemical potential,''
Phys. Rev. D \textbf{66} (2002), 074507
doi:10.1103/PhysRevD.66.074507
[arXiv:hep-lat/0204010 [hep-lat]].
%680 citations counted in INSPIRE as of 27 Oct 2021

%\cite{Gavai:2008zr}
\bibitem{Gavai:2008zr}
R.~V.~Gavai and S.~Gupta,
%``QCD at finite chemical potential with six time slices,''
Phys. Rev. D \textbf{78} (2008), 114503
doi:10.1103/PhysRevD.78.114503
[arXiv:0806.2233 [hep-lat]].
%270 citations counted in INSPIRE as of 28 Nov 2021

%\cite{Bellwied:2015lba}
\bibitem{Bellwied:2015lba}
R.~Bellwied, S.~Borsanyi, Z.~Fodor, S.~D.~Katz, A.~Pasztor, C.~Ratti and K.~K.~Szabo,
%``Fluctuations and correlations in high temperature QCD,''
Phys. Rev. D \textbf{92} (2015) no.11, 114505
doi:10.1103/PhysRevD.92.114505
[arXiv:1507.04627 [hep-lat]].
%163 citations counted in INSPIRE as of 28 Nov 2021

%\cite{Bazavov:2017dus}
\bibitem{Bazavov:2017dus}
A.~Bazavov, H.~T.~Ding, P.~Hegde, O.~Kaczmarek, F.~Karsch, E.~Laermann, Y.~Maezawa, S.~Mukherjee, H.~Ohno and P.~Petreczky, \textit{et al.}
%``The QCD Equation of State to $\mathcal{O}(\mu_B^6)$ from Lattice QCD,''
Phys. Rev. D \textbf{95} (2017) no.5, 054504
doi:10.1103/PhysRevD.95.054504
[arXiv:1701.04325 [hep-lat]].
%303 citations counted in INSPIRE as of 27 Oct 2021

%\cite{HotQCD:2018pds}
\bibitem{HotQCD:2018pds}
A.~Bazavov \textit{et al.} [HotQCD],
%``Chiral crossover in QCD at zero and non-zero chemical potentials,''
Phys. Lett. B \textbf{795} (2019), 15-21
doi:10.1016/j.physletb.2019.05.013
[arXiv:1812.08235 [hep-lat]].
%273 citations counted in INSPIRE as of 28 Nov 2021

%\cite{Giordano:2019slo}
\bibitem{Giordano:2019slo}
M.~Giordano and A.~P\'asztor,
%``Reliable estimation of the radius of convergence in finite density QCD,''
Phys. Rev. D \textbf{99} (2019) no.11, 114510
doi:10.1103/PhysRevD.99.114510
[arXiv:1904.01974 [hep-lat]].
%16 citations counted in INSPIRE as of 27 Oct 2021


%\cite{deForcrand:2002hgr}
\bibitem{deForcrand:2002hgr}
P.~de Forcrand and O.~Philipsen,
%``The QCD phase diagram for small densities from imaginary chemical potential,''
Nucl. Phys. B \textbf{642} (2002), 290-306
doi:10.1016/S0550-3213(02)00626-0
[arXiv:hep-lat/0205016 [hep-lat]].
%779 citations counted in INSPIRE as of 26 Nov 2021

%\cite{DElia:2002tig}
\bibitem{DElia:2002tig}
M.~D'Elia and M.~P.~Lombardo,
%``Finite density QCD via imaginary chemical potential,''
Phys. Rev. D \textbf{67} (2003), 014505
doi:10.1103/PhysRevD.67.014505
[arXiv:hep-lat/0209146 [hep-lat]].
%606 citations counted in INSPIRE as of 27 Nov 2021

%\cite{Cea:2015cya}
\bibitem{Cea:2015cya}
P.~Cea, L.~Cosmai and A.~Papa,
%``Critical line of 2+1 flavor QCD: Toward the continuum limit,''
Phys. Rev. D \textbf{93} (2016) no.1, 014507
doi:10.1103/PhysRevD.93.014507
[arXiv:1508.07599 [hep-lat]].
%79 citations counted in INSPIRE as of 27 Nov 2021

%\cite{DElia:2016jqh}
\bibitem{DElia:2016jqh}
M.~D'Elia, G.~Gagliardi and F.~Sanfilippo,
%``Higher order quark number fluctuations via imaginary chemical potentials in $N_f=2+1$ QCD,''
Phys. Rev. D \textbf{95} (2017) no.9, 094503
doi:10.1103/PhysRevD.95.094503
[arXiv:1611.08285 [hep-lat]].
%81 citations counted in INSPIRE as of 27 Nov 2021

%\cite{Bonati:2018nut}
\bibitem{Bonati:2018nut}
C.~Bonati, M.~D'Elia, F.~Negro, F.~Sanfilippo and K.~Zambello,
%``Curvature of the pseudocritical line in QCD: Taylor expansion matches analytic continuation,''
Phys. Rev. D \textbf{98} (2018) no.5, 054510
doi:10.1103/PhysRevD.98.054510
[arXiv:1805.02960 [hep-lat]].
%65 citations counted in INSPIRE as of 27 Nov 2021

%\cite{Bellwied:2019pxh}
\bibitem{Bellwied:2019pxh}
R.~Bellwied, S.~Borsanyi, Z.~Fodor, J.~N.~Guenther, J.~Noronha-Hostler, P.~Parotto, A.~Pasztor, C.~Ratti and J.~M.~Stafford,
%``Off-diagonal correlators of conserved charges from lattice QCD and how to relate them to experiment,''
Phys. Rev. D \textbf{101} (2020) no.3, 034506
doi:10.1103/PhysRevD.101.034506
[arXiv:1910.14592 [hep-lat]].
%33 citations counted in INSPIRE as of 26 Nov 2021

%\cite{Borsanyi:2020fev}
\bibitem{Borsanyi:2020fev}
S.~Borsanyi, Z.~Fodor, J.~N.~Guenther, R.~Kara, S.~D.~Katz, P.~Parotto, A.~Pasztor, C.~Ratti and K.~K.~Szabo,
%``QCD Crossover at Finite Chemical Potential from Lattice Simulations,''
Phys. Rev. Lett. \textbf{125} (2020) no.5, 052001
doi:10.1103/PhysRevLett.125.052001
[arXiv:2002.02821 [hep-lat]].
%87 citations counted in INSPIRE as of 27 Oct 2021

%\cite{Pasztor:2020dur}
\bibitem{Pasztor:2020dur}
A.~P\'asztor, Z.~Sz\'ep and G.~Mark\'o,
%``Apparent convergence of Pad\'e approximants for the crossover line in finite density QCD,''
Phys. Rev. D \textbf{103} (2021) no.3, 034511
doi:10.1103/PhysRevD.103.034511
[arXiv:2010.00394 [hep-lat]].
%6 citations counted in INSPIRE as of 26 Nov 2021

%\cite{Bellwied:2021nrt}
\bibitem{Bellwied:2021nrt}
R.~Bellwied, C.~Ratti, S.~Borsanyi, P.~Parotto, Z.~Fodor, J.~N.~Guenther, S.~D.~Katz, A.~Pasztor, D.~Pesznyak and K.~K.~Szabo,
%``Corrections to the hadron resonance gas from lattice QCD and their effect on fluctuation-ratios at finite density,''
Phys. Rev. D \textbf{104} (2021) no.9, 094508
doi:10.1103/PhysRevD.104.094508
[arXiv:2102.06625 [hep-lat]].
%3 citations counted in INSPIRE as of 26 Nov 2021

%\cite{Borsanyi:2021sxv}
\bibitem{Borsanyi:2021sxv}
S.~Bors\'anyi, Z.~Fodor, J.~N.~Guenther, R.~Kara, S.~D.~Katz, P.~Parotto, A.~P\'asztor, C.~Ratti and K.~K.~Szab\'o,
%``Lattice QCD equation of state at finite chemical potential from an alternative expansion scheme,''
Phys. Rev. Lett. \textbf{126} (2021) no.23, 232001
doi:10.1103/PhysRevLett.126.232001
[arXiv:2102.06660 [hep-lat]].
%16 citations counted in INSPIRE as of 27 Oct 2021


%\cite{Parisi:1983mgm}
\bibitem{Parisi:1983mgm}
G.~Parisi,
%``ON COMPLEX PROBABILITIES,''
Phys. Lett. B \textbf{131} (1983), 393-395
doi:10.1016/0370-2693(83)90525-7
%303 citations counted in INSPIRE as of 28 Nov 2021

%\cite{Aarts:2009uq}
\bibitem{Aarts:2009uq}
G.~Aarts, E.~Seiler and I.~O.~Stamatescu,
%``The Complex Langevin method: When can it be trusted?,''
Phys. Rev. D \textbf{81} (2010), 054508
doi:10.1103/PhysRevD.81.054508
[arXiv:0912.3360 [hep-lat]].
%212 citations counted in INSPIRE as of 28 Nov 2021

%\cite{Sexty:2013ica}
\bibitem{Sexty:2013ica}
D.~Sexty,
%``Simulating full QCD at nonzero density using the complex Langevin equation,''
Phys. Lett. B \textbf{729} (2014), 108-111
doi:10.1016/j.physletb.2014.01.019
[arXiv:1307.7748 [hep-lat]].
%177 citations counted in INSPIRE as of 27 Oct 2021


%\cite{deForcrand:2002pa}
\bibitem{deForcrand:2002pa}
P.~de Forcrand, S.~Kim and T.~Takaishi,
%``QCD simulations at small chemical potential,''
Nucl. Phys. B Proc. Suppl. \textbf{119} (2003), 541-543
doi:10.1016/S0920-5632(03)80451-6
[arXiv:hep-lat/0209126 [hep-lat]].
%29 citations counted in INSPIRE as of 27 Oct 2021

%\cite{Alexandru:2005ix}
\bibitem{Alexandru:2005ix}
A.~Alexandru, M.~Faber, I.~Horvath and K.~F.~Liu,
%``Lattice QCD at finite density via a new canonical approach,''
Phys. Rev. D \textbf{72} (2005), 114513
doi:10.1103/PhysRevD.72.114513
[arXiv:hep-lat/0507020 [hep-lat]].
%125 citations counted in INSPIRE as of 27 Oct 2021

%\cite{Borsanyi:2010bp}
\bibitem{Borsanyi:2010bp}
S.~Borsanyi \textit{et al.} [Wuppertal-Budapest],
%``Is there still any $T_c$ mystery in lattice QCD? Results with physical masses in the continuum limit III,''
JHEP \textbf{09} (2010), 073
doi:10.1007/JHEP09(2010)073
[arXiv:1005.3508 [hep-lat]].
%1000 citations counted in INSPIRE as of 28 Nov 2021

\end{thebibliography}
\end{document}